# MECHANISMS INFLUENCING THE FORMATION OF THE CARBONIZED LAYER RELIEF ON THE POLYMER SURFACE AFTER THE ION-PLASMA TREATMENT MODELING

## A.Yu. Beliaev [1,2], A.L. Svistkov [1,2]

[1] Laboratory of Micromechanics of Structural and Heterogeneous Mediums, Institute of Continuous Media Mechanics, Ural Branch of Russian Academy of Sciences, 1 Acad. Korolev Street, 614013, Perm, Russia, e-mail: belyaev@icmm.ru
[2] Department of Continuous Media Mechanics and Computer Technology, Perm State National Research University, 15 Bukirev Street, 614990, Perm, Russia

**Abstract.** The work is devoted to modeling of the processes occurring in ion-plasma treatment of polyurethane and forming on its surface a special carbonized layer. The mechanisms affecting the formation of wave-like relief in the carbonized layer are examined. The boundary value problem for the heat equation for modeling the effect of temperature component was solved. Using the finite element method the deformation of the polyurethane sample during ion-plasma treatment was calculated. On the basis of calculations conclusion about the impact of these factors on the formation of wave-like relief in the carbonized layer was made.

## Introduction

The human body is a very complex system. Any interference to this system can lead to undesirable consequences. To solve many problems in modern medicine (especially implantology) a detailed study of the materials implemented to a human organism is required. The question is not only about the biological but also the mechanical compatibility of the implant with the body. To solve such problems it is necessary to resort both to experimental methods for studying materials [1] and to methods of mathematical modeling. Mathematical models will help to understand the processes taking place in the system under consideration [2,3]. Together with mechanical tests this will enable to develop and optimize the type of implant [4,5] which is necessary to solve the problem. In this paper we consider the simulation of the processes occurring during ion-plasma treatment of polyurethane and the formation of a special carbonized layer on its surface. The information obtained will help in the development of polyurethane biocompatible implants.

Ion-plasma treatment of materials is of great interest in many fields. For instance it is the treatment of polymers for the creation of implants and medical equipment [6-8]. Ion-plasma treatment of the polymer leads to the formation of a special carbonized layer on its surface (Figure 1) which has essential properties for medical use [9-10]. A protein which acts as an intermediate link between the implant and the body and prevents rejection, inflammation of surrounding tissues and other unfavorable situations, can be planted on this layer [11-12]. In addition to physical and chemical properties the important characteristic of the layer is the relief and roughness of the treated surface. First the protein and living cells are Anton Yu. Beliaev, Junior researcher of Laboratory of Micromechanics of Structural and Heterogeneous Mediums, Perm
Aleksandr L. Svistkov, PhD, Head of Laboratory of micromechanics of Structural-heterogeneous Mediums, Perm

sensitive to the relief of the surface on which they are attached. Secondly it is undesirable to allow cracking of the carbonized layer since this can lead to the initiation and further development of microcracks on the surface of the implant. In this regard, it is important to analyze the factors that influence the formation of the relief of the carbonized layer.

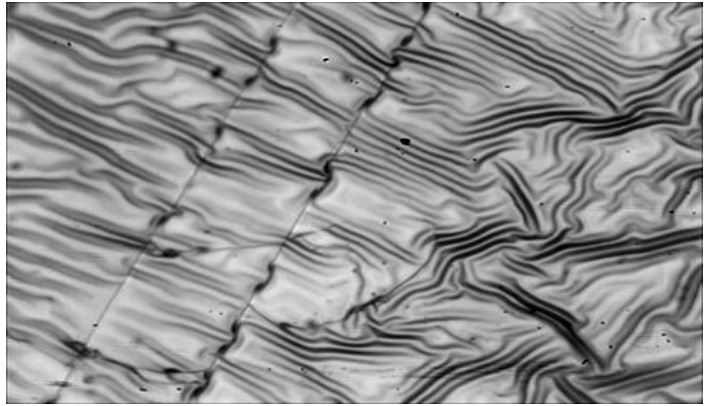

Fig. 1 Surface of polyurethane after ion-plasma treatment

The carbonized layer is formed due to the restructuring of the molecular structure of polyurethane, links in which break under the action of ions. In fact the carbonized layer is a thin plate on the surface of a more massive polyurethane substrate. Wave formation of the layer can be explained by the fact that after ion-plasma treatment oxygen is actively integrates into the polymer network [13]. There is a swelling of the surface layer which leads to a loss of stability and the formation of waves as a result. However in the opinion of the authors there are other mechanisms of loss of stability having a mechanical nature. These mechanisms are discussed in this paper.

## FORMULATION OF THE PROBLEM

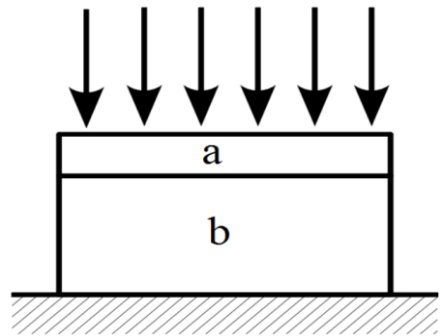

Fig. 2 The process of a polyurethane sample treatment by plasma calculation scheme

Figure 2 shows the calculation scheme: the ion flux acts on the body (a) located on the elastic base (b). In this scheme, the body (a) is a carbonized layer formed on the surface of the polyurethane (b) during the treatment.

Let us write down the first law of thermodynamics. The rate of change in the kinetic energy, elastic energy and the amount of heat accumulated by the system depends on the speed of the ion transported by the flow of kinetic energy and on the capacity of the work. This is formulated using a mathematical equation:

$$\left[\frac{M\vartheta^2}{2} + \frac{k\Delta l^2}{2l_0^2} + c\Delta T\right]' = NS\frac{m\vartheta_0^2}{2} + \frac{k\Delta l}{l_0}\vartheta \tag{1}$$

where M is the mass of the body located on the elastic base, $\upsilon$ is the load speed at the current time, k is the stiffness of the base, $\Delta l$ is the elongation of the base during deformation, $l_0$ is the initial height of the base, c is the specific heat of the system, $\Delta T$ is the temperature change, N is the number of ions incident per area unit, per time unit, S is the area of the ions action, m is the mass of one nitrogen ion, $\upsilon_0$ is the rate at which the nitrogen ion approaches the polyurethane surface. The value of $\upsilon_0$ is known. A hatch at the closing parenthesis denotes differentiation with respect to time.

Applying the Galilean transformation we obtain the following equation:

$$\left[\frac{M(\vartheta+\vartheta^*)^2}{2} + \frac{k\Delta l^2}{2l_0^2} + c\Delta T\right]' = NS\frac{m(\vartheta_0+\vartheta^*)^2}{2} + \frac{k\Delta l}{l_0}(\vartheta+\vartheta^*) \tag{2}$$

where $\upsilon^*=const$ − movement speed of the observer. The conservation law of energy (1) must be satisfied for an observer moving with any constant velocity $\vartheta^*$. It means that it is possible to group the terms in (2) in such way that we get a quadratic equation with respect to the degrees of $\vartheta^*$.

$$(\ldots)\vartheta^{*2} + (\ldots)\vartheta^* + (\ldots) = 0$$

This equality can be satisfied for any value of the velocity $\vartheta^*$ only in the case when the expressions in parentheses in front of the powers of $\vartheta^*$ equal zero. We obtain 3 independent equations:

$$\dot{M} = NSm \tag{3}$$

$$M\dot{\vartheta} = NSm\vartheta_0 - k\frac{\Delta l}{l_0} \tag{4}$$

$$\left[\frac{M\vartheta^2}{2} + \frac{k\Delta l^2}{2l_0^2} + c\Delta T\right]' = NSm\frac{\vartheta_0^2}{2} + \frac{k\Delta l}{l_0}\vartheta \tag{5}$$

Equation (3) is the equation of mass change. The mass of the system is increased by an amount equal to the mass of all ions that have reached the target in a certain time.

Equation (4) is the equation of motion of the system. Deformation of the treated material occurs under the influence of ion pressure. The resulting equation describes the displacement of the treated surface under the influence of ion pressure. The mechanism of the appearance of waves under ion pressure and the solution of this equation will be discussed in detail later.

Equation (5) is the heat equation. Getting into the material that is being treated the ion spends part of the energy on the destruction of molecular bonds due to which the carbonized layer is formed. However most of the energy simply dissipates into the heat. The temperature mechanism of waves appearance and the solution of this equation will be considered below.

## IONS PRESSURE MECHANISM OF WAVE SURFACE APPEARANCE

First we consider the mechanism of the appearance of a wavelike surface in polyurethane in the process of ion-plasma treatment. It is related to the ion pressure which causes deformation in the material. As was mentioned in the previous chapter the ion energy is quite large. Each ion in a collision with a surface acts on it with a force that can be calculated from the kinematic relationships of mechanics. Taking into account that the radiation dose is $10^{14}$-$10^{16}$ ions / cm2, it can be said that a constant pressure acts on the surface during the treatment. If the pressure is large enough it can cause elastic deformation in the material. The layer is formed on deformed material. After the treatment process is finished the material returns to its original, unstrained state. The carbonized layer has a thickness of about 50 nm, the thickness of the polyurethane plate is 200 μm. The difference in thickness is so great that at any ratio of rigidity deformation of the plate will cause deformation of the layer. Thus after the flow of ions has stopped, returning of the substrate to its original state will cause deformation in the formed layer; the layer was formed on a deformed plate and this state will be the initial one for it. Figure 3 shows the expected scheme of the appearance of waves on the surface of the treated material.

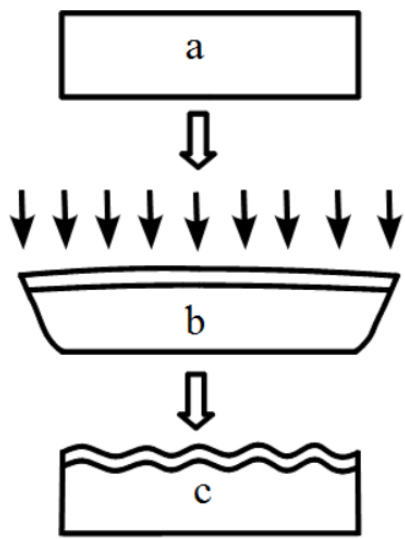

Fig. 3 a-initial state of the sample, b-process of ion-plasma treatment, c-transition to the equilibrium state after ion-plasma treatment of the material

The result of simulation of a layer deformation by a substrate is shown in Fig. 4. The calculation showed that when the polyurethane is unloaded, the carbonized layer deforms and this leads to the formation of a wavy relief, in case when the deformation of the sample exceeds a certain value.

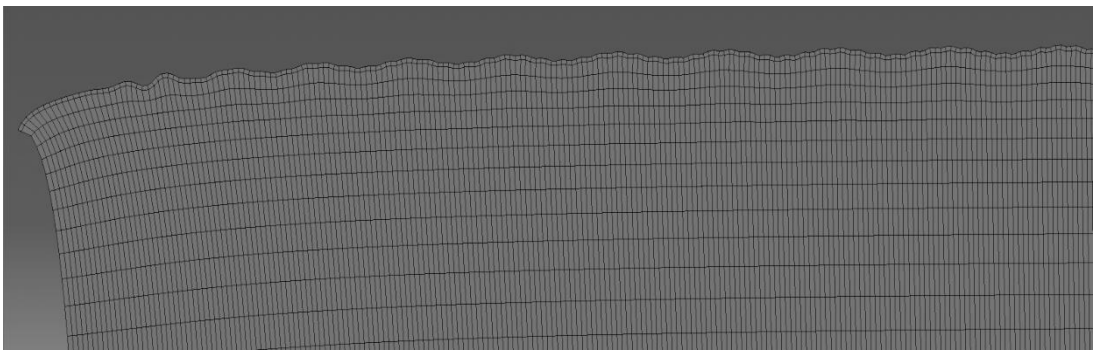

Fig. 4 Formation of a wave-like relief in a carbonized layer after depressurization from a polyurethane substrate

The solution of equation (4) will gives the dependence l (t), i.e. displacement of the layer on the elastic substrate under the action of ion pressure. In other words we get an expression for moving of the surface during the treatment. Knowing the displacement you can get the values of the deformation which will predict the folds appearance.

We can get the solution of equation (4) using the initial conditions:

$$l|_{t=0} = l_0$$

$$l'|_{t=0} = 0$$

The first condition means that at time t=0 the layer is in the initial position $l_0$ (the layer thickness can be neglected, since it is negligible compared to the thickness of the polyurethane being treated). Solving equation (4) we obtain an expression for l(t) in the form:

$$l(t) = l_0 - \frac{l_0}{k} NSm\vartheta_0 \left[ 1 + \cos \sqrt{\frac{k}{Ml_0}} t \right] \qquad (6)$$

The analysis of expression (6) has shown that the treated surface is moving according to a harmonic law. This is the result of the fact that we considered the polymer as an elastic medium and did not take into account its dissipative properties. The solution is a constant deformation of the material and an oscillatory process near this state. The magnitude of the strain is inversely proportional to the stiffness of the system, proportional to the ion flux density and ion velocity. The ion velocity in turn depends on the energy. Thus we can distinguish two basic parameters that affect the amount of deformation: ion energy and polyurethane stiffness.

In this calculation, the surface displacement values reach 0.2 mm. Such displacements are not enough to cause deformations necessary for the formation of folds.

This mechanism with low probability can have a significant influence on the formation of wave-like relief. However in rare cases if it is planned to use soft polyurethane for the implant it is necessary to consider this mechanism more carefully.

In connection with the small significance of this mechanism we consider the following mechanism.

### THE WAVE- SURFACE APPEARING TEMPERATURE MECHANISM

The second of the proposed mechanisms for the appearance of waves on the surface of polyurethane is associated with an increasing of the temperature during ion-plasma treatment. When the ion, which has received the kinetic energy, reaches the surface of the polyurethane it spends part of this energy on breaking the bonds of atoms in the polymer molecules. Due to this the surface is modified and a carbonized layer is formed. However the energy expended on breaking bonds is small in comparison with the energy of the ion. The remaining kinetic energy is dissipated into heat. As the temperature increases the polyurethane expands. The carbonized layer is formed already on a material deformed by thermal expansion. When the treatment process ends the material and the carbonized layer cools. The coefficient of linear thermal expansion of polyurethane is $5 \cdot 10^{-5}$ -$7 \cdot 10^{-5}$ 1/K which is several times greater than of carbon ($8 \cdot 10^{-6}$ -$2 \cdot 10^{-5}$ 1/K). This means that the temperature deformations of the polyurethane and the carbon layer can differ significantly. In the carbon layer temperature strain will be less than in polyurethane which will cause the appearance of waves. This assumption seems convincing provided that the ions heat the polyurethane to sufficiently high temperatures. In this regard it is necessary to estimate the temperature of the material during the ion-plasma treatment.

Equation (5) gives the solution $\Delta T$ (t). Thus we obtain an average value of the temperature over the thickness of the sample depending on the time. This solution does not provide information on the spatial distribution of temperature during treatment. Due to the fact that the polyurethane has a low thermal conductivity the distribution of temperature in the thickness of the sample is nontrivial. Let us consider the problem in another formulation which makes it possible to obtain a spatial temperature distribution over the thickness of the sample being treated.

Consider the case when the polyurethane is treated with nitrogen ions with the energy of 0.1 keV. The fluence is $2*10^{16}$ ions/cm $^2$ and is achieved in 47 seconds. For a rough estimate of the heating let us take the total energy of the ions (which under the conditions mentioned above act on a sample area of 1 cm$^2$ and a thickness of 0.2 mm, density of 1100 kg / m$^3$) and calculate the temperature change according to formula (7).

$$Q = \mu m \Delta T \tag{7}$$

where Q is the energy obtained as a result of ions deceleration when they meet the polyurethane sample, $\mu$ is the specific heat of polyurethane, $\Delta T$ is the temperature change. We assume that all the kinetic energy of the ions is dissipated in to the heat. This estimation is excess. In fact some of the energy is spent on the destruction of the polymer and additional release can occur during the formation of the carbon layer.

The estimate does not take into account the heat sink from the material. In reality the sample is located on a massive metal holder which has a large heat capacity and thermal conductivity than polyurethane. Thus in order for the assessment to be more adequate it is necessary to take into account the heat flow through the surface in contact with the holder. For this we set the following boundary-value problem:

$$\rho c \frac{\partial T}{\partial t} = \mu \frac{\delta^2 T}{\delta x^2} + s(x, t) \tag{8}$$

where T is the temperature, $\rho$ is the density, c is the specific heat, $\mu$ is the thermal conductivity of the polyurethane, and s is the specific power of the source. Formula (2) is the heat equation. The problem is posed as one-dimensional since the field of sources is uniform and the temperature gradient is present only in the direction of the thickness of the sample. Ions are understood as the source of heat in the model. It is believed that distribution of ions are uniform over the volume of the layer of 5 nm (the thickness of the layer is calculated using the software package SRIM). The specific power of a source is understood as the energy released per unit of time per volume unit. In this case the ion energy is 0.1 keV.

Because of the smallness of the sources we will consider a continuous, homogeneous, 5 nm thick field of heat sources with a total thickness of the polyurethane plate of 200 μm. The source operates for 47 seconds (duration of sample treatment with nitrogen ions) and then turns off. The rest of the time there is a free redistribution of heat in the thickness of the plate and its exit through the lower boundary. In this model the energy flow through the holder is taken into account which gives us the condition on the right boundary. The specific heat of the steel is 4620 J / (kg · K), and the thermal conductivity coefficient is 47 W / (m·K). Similar characteristics of the polyurethane are 1100 J / (kg · K) and 0.041 W / (m · K). In this regard, the steel holder will absorb heat from the polyurethane at a high rate and provide a sink of energy. The initial temperature is taken zero because only relative change in temperature is calculated. Thus the initial and boundary conditions will have the form:

$$T(x, 0) = 0$$

$$\frac{\partial T}{\partial x}(0, t) = 0 \qquad (9)$$
$$U(200, t) = 0$$

Having a sink at the lower boundary we should not have received high temperatures but taking into account the low thermal conductivity of the polyurethane it is possible to obtain a situation where the energy does not have time to pass through the entire thickness of the plate to runoff fast enough and it will lead to a strong heating of the upper boundary of the plate. The solution showed that due to the small thickness of the plate the process of transition to a stationary state occurs instantaneously and a balance is rapidly established between the flow and the sink so that there is no strong heating.

Figure 5 shows the temperature dependence on time at the boundary on which the sources act. It can be noted that the output to the stationary state occurs almost instantaneously. Also after switching off the source the temperature quickly returns to the initial temperature. During ion-plasma treatment the temperature rises to a maximum of 0.6 degrees. An important factor here is the sink of the heat on the opposite boundary. Figure 6 shows the temperature distribution in the absence of a heat sink under the same conditions. Under these conditions the temperature rises more than 20 degrees. A similar effect is observed when processing massive samples. Figure 7 shows the temperature distribution at the boundary of the processed sample with 2 cm thickness. As can be seen as the thickness of the sample increases the treated surface heats up more.

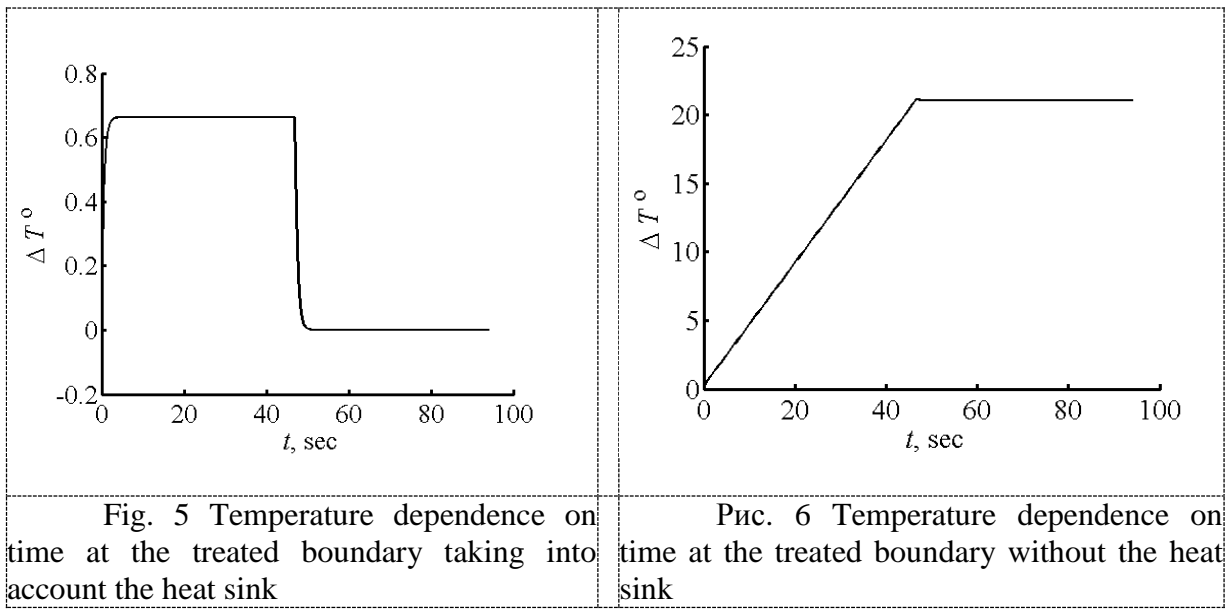

Fig. 5 Temperature dependence on time at the treated boundary taking into account the heat sink

Рис. 6 Temperature dependence on time at the treated boundary without the heat sink

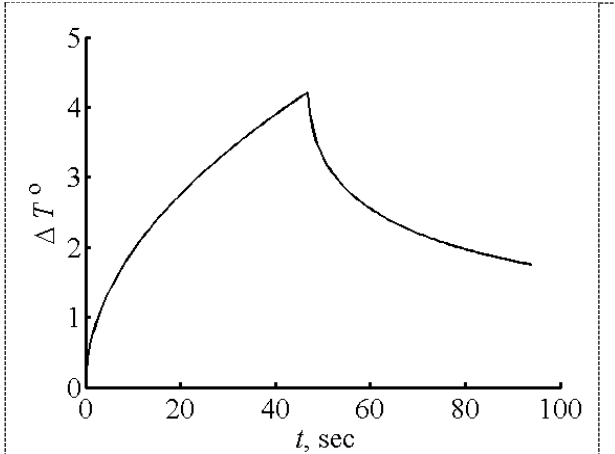
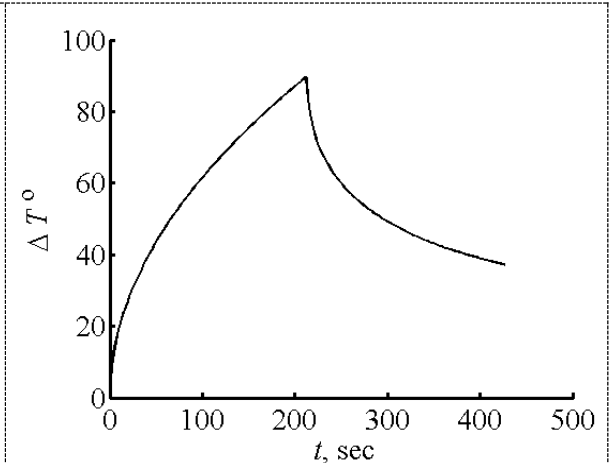

| Fig. 7 Temperature dependence on time on the treated boundary of a thick sample (2 cm) | Fig. 8 Temperature dependence on time on the treated boundary of a thick sample (2 cm) at an ion energy of 1keV |

This effect is connected with low thermal conductivity of the polyurethane. In fact the sample itself is a heat isolator. The time for which heat reaches the boundary with leak increases and during this time the surface heats up to a higher temperature. In a process of treatment of a thick sample with ion energy of 1 keV (Figure 8) the temperature change reaches more than 80 degrees. If sufficient heat sink is not provided the surface of the sample can be heated to the critical temperatures at which the thermal degradation of the polymer occurs.

Thus this mechanism depends on the ion energy, massiveness of the treated object and the conditions of heat sink from the sample. It is especially relevant for sufficiently thick samples and high ion energies.

## CONCLUSIONS

The calculations have shown that the temperature mechanism affects the formation of the relief of the carbonized layer. Despite the fact that the swelling of the surface layer (due to the incorporation of oxygen atoms) has a determining effect the temperature mechanism also contributes to the formation of the relief. In cases of treatment at high ion energies and processing of massive samples the temperature mechanism can be decisive. The physical and chemical properties of the surface depend on the energy. By varying the magnitude of the energy it is possible to achieve different surface roughness, the length and height of the formed waves. It is important to note that at high ion energies one can encounter the problem of strong heating of the treated object and as a consequence thermal destruction of the polymer.

Improper selection of parameters of ion-plasma treatment can lead to the formation of such a relief that the interaction of living cells with which will be unsatisfactory in the problem of biocompatibility achieving. Received models allow getting conclusions about the influence of the considered mechanisms on the formation of the relief.

The mathematical model of the swelling of the formed carbonized layer is not considered in the article in view of the complexity of the problem of determining constants entering into equations. The purpose of this paper was to find the answer to the question whether it is possible to explain the formation of the wave-like surface of a sample with the help of other mechanisms and to determine the conditions for it. The calculations have shown that in some cases the cause of the observed relief of the surface can be explained by the temperature effects that occur during the treatment of the material.

*This work was supported by the Russian Fund for Basic Research (grant № 16-08-00910_a, 17-48-590057_r_a).*